%% file: main.tex
\documentclass{article}[12pt]
\usepackage{graphicx} 
\usepackage[margin=1in]{geometry}
\usepackage{subcaption}
\usepackage{amsmath}
\usepackage{bm} 
\usepackage{adjustbox}
\usepackage{amssymb}
\usepackage[affil-it]{authblk}

\def\edgecolor{rgb:blue,4;red,1;green,4;black,3}

\usepackage{tikz}
\usetikzlibrary{positioning,shapes}
\usetikzlibrary{3d} 
\usetikzlibrary{shapes.geometric, quotes, arrows.meta, decorations.pathreplacing}
\usepackage{import}
\input{init}

\usepackage{natbib}

\title{Using Images as Covariates: Measuring Curb Appeal with Deep Learning\thanks{ Thanks to audience members at the 17th International Conference on Computational and Financial Econometrics and Carleton University. Webb thanks the Social
Sciences and Humanities Research Council of Canada (SSHRC grant
435-2021-0396) for financial support. We thank chatGPT for valuable coding and tikz assistance. }}

\author[1]{Ardyn Nordstrom}
\author[2]{Morgan Nordstrom}
\author[3]{Matthew D. Webb}

\affil[1]{Carleton University, SPPA}
\affil[2]{Carleton University, Department of Geography}
\affil[3]{Carleton University, Department of Economics}

\date{\today}

\begin{document}

\maketitle

\begin{abstract}
    \noindent This paper details an innovative methodology to integrate image data into traditional econometric models. Motivated by forecasting sales prices for residential real estate, we harness the power of deep learning to add “information” contained in images as covariates. Specifically, images of homes were categorized and encoded using an ensemble of image classifiers (ResNet-50, VGG16, MobileNet, and Inception V3). Unique features presented within each image were further encoded through panoptic segmentation. Forecasts from a neural network trained on the encoded data results in improved out-of-sample predictive power. We also combine these image-based forecasts with standard hedonic real estate property and location characteristics, resulting in a unified dataset. We show that image-based forecasts increase the accuracy of hedonic forecasts when encoded features are regarded as additional covariates. We also attempt to “explain” which covariates the image-based forecasts are most highly correlated with. The study exemplifies the benefits of interdisciplinary methodologies, merging machine learning and econometrics to harness untapped data sources for more accurate forecasting.

\end{abstract}

\section{Introduction}

The presence of unobservables is one of the most fundamental challenges in empirical economics research. Deep learning methods can help researchers across disciplines capture some of these unobservables by accessing information embedded in images, videos, and other types of unstructured data. By combining multiple layers of processing, deep learning makes it possible to incorporate this data in standard econometric approaches. These methods have been widely adopted in other disciplines like medicine \citep[e.g.,][]{ker2017deep} and biology \citep[e.g.,][]{moen2019deep} where image-based data is more common. However, new sources of unstructured data such as satellite images, traffic photos, and photos from social media are now more widely available, presenting new opportunities for disciplines like economics. This raises the question: how can this type of data be used to answer economic questions?
In this paper, we attempt to answer this by showing how employing convolutional neural networks can convert unstructured data in images into usable information for economic research. 

Given the importance of visual features of housing in purchasing decisions, we assess whether image data can improve housing price predictions that come from standard hedonic analyses. Hedonic pricing models offer a relatively straightforward approach to estimate housing prices using observable characteristics about a house and its location and are widely used in economics. However, the standard hedonic pricing model can overlook important but unobservable attributes that influence how houses are priced. 

Consider two hypothetical houses that are similar in terms of their bedrooms, bathrooms, square footage, lot size, features, and neighborhood characteristics. This type of information is traditionally captured in structured real estate data. 
While two houses may be similar across these features, many other characteristics influence the perceived worth of each property. These kinds of differences are observable when potential buyers look at photos of, or visit, the property but are not captured in structured housing data. 

To capture some of these unobservables we use deep learning methods, a subset of machine learning, that combine multiple layers of neural networks to process data. Unlike machine learning methods that economists are more familiar with, deep learning methods do not rely on input data to have a specific representation or structure \citep{goodfellow2016deep}. 
Instead, deep learning can be used to detect the representations that exist within this unstructured data. In the case of images, each layer of a convolutional neural network identifies increasingly comprehensive patterns within the pixels that make up the image until later layers can identify objects and other features in the data \citep{lecun2015deep, goodfellow2016deep}.  

When convolutional neural networks are used to identify features in unstructured data, these are typically referred to as ``architectures''. Many pre-trained architectures have been used in the literature. For example, the recent \citet{LM_2024} paper uses ``ResNet50,'' a widely-used architecture that was trained on the ImageNet dataset containing over ten million photos \citep{he2016deep}. However, there are many other trained architectures, which are each able to identify different features within a photo. For applications that exclusively use pre-trained architectures to identify features in images, these parts of the architectures are often referred to as ``encoders''. While most applications to date \citep[e.g.][]{LM_2024, law2019take, yeh2020using} have used one encoder to identify the features in images, we use multiple encoders. In a recent working paper, \citet{compiani2023demand} uses multiple encoders to measure the similarity of photos of consumer products to measure price elasticities. To our knowledge, we are the first to use multiple encoders to identify features for housing predictions based on unstructured data. We find the use of multiple encoders improves the out-of-sample predictive power of our models. 

Our work builds on the seminal work by \citet{yeh2020using}, who use deep learning based on the ResNet18 encoder (a predecessor to ResNet50) to predict wealth levels of areas using images captured by satellites. \citet{yeh2020using} show that deep learning can be used to measure common economic indicators from publicly available satellite images. Since this data is frequently missing in the African contexts they focus on, this work highlights the utility of using deep learning to capture traditionally unobserved data. Others have since demonstrated how using the unobservables detected using deep learning can improve economic predictions and evaluation research. The use of satellite imagery data is now widely used in economics research to use information about ground conditions from satellite imagery \citep{donaldson2016view, takahashi2017coffee, asher2020rural}. For example, a recent related working paper by \citet{han2024cool} shows how satellite imagery data can be used to detect tree cover loss in Toronto. They use the information from these images to estimate the impact that tree cover loss has on housing prices, however, they do not rely on deep learning to identify features in the satellite images.

Using a similar approach to \citet{yeh2020using}, a recent working paper by \citet{huang2021using} uses the ResNet50 architecture to estimate the quality of housing in areas in Kenya that had recently been exposed to a cash transfer program. They find that using satellite images alone can measure the impact of this kind of intervention. 
\citet{LM_2024} apply this approach to a different context to use deep learning in questions related to criminal justice. While they develop a framework for how deep learning can be used to identify new hypotheses from unstructured data, they apply their procedure to make predictions about judges' sentencing decisions. By using the encoded features (again using ResNet50) from mugshot images, along with structured information about a defendant, they can explain how a defendant's appearance influences a judge's sentencing decisions. These applications both demonstrate how the features from single encoders can make image data useful in economic research.

Given the importance of visual aspects of properties in determining real estate values, housing has been used as an application in the computer science literature to test how deep learning can be used to conduct visual content analysis and scene understanding. \citet{law2019take} develop and train two separate architectures to identify features in Google Street View images and satellite images, respectively, in London, UK. These features are used to predict a property's desirability, which they use to predict housing prices. Like in our study, they find that information about a location's desirability can improve the accuracy of predictive house price models. However, unlike \citet{law2019take}, we show that these gains can be achieved using multiple well-established, pre-trained encoders that economists are becoming more familiar with, using only a single exterior photo from a property. 
We also extend the work of \citet{zhao2019deep}, who used deep learning models to process interior and exterior photos of houses from listings in Australia to come up with a measure of how aesthetically pleasing the photos were. This measure was then used as an input in a model that used multiple machine learning methods to predict housing prices.

This discussion provides an outline for how our current work differs from and contributes to the existing literature. The first contribution is related to our methodology. Existing research that uses deep learning models has so far primarily focused on using one or two prediction methods. In \citet{LM_2024}, predictions come from a convolutional neural network. In \citet{law2019take}, predictions come from both ordinary least squares and a convoluted model that augments a linear model with the predictions of desirability from a neural network. This was based on work by \citet{peterson2009neural}, \citet{ahmed2016house}, which compare the results of predictions from standard linear hedonic price models to predictions from neural networks. We take a broader approach to compare the predictions from each of the models considered in the literature: ordinary least squares, neural networks, and a convoluted model that uses predictions from a neural network in ordinary least squares.

The second contribution is related to how novel methods can be used to capture unobservables in unstructured data. For example, \citet{voice_2023}, \citet{face_2023}, and \citet{words_2023} have used information from vocal tone, video, and text data to capture unobservables in press conferences from the Federal Reserve, respectively. 
Similar to how we are trying to incorporate the visual characteristics of a property that contain information about its value, these papers recognize that tone and body language contain information about the message being communicated.\footnote{
Specifically, \cite{words_2023} uses video, audio, and text from the Federal Reserve Chair's meetings to assess the emotions of the meetings. They find that both stock prices and volatility 
respond to the Fed Chair's emotional indices. 
\cite{voice_2023} analyze Fed Chair's vocal tone during (Federal Open Market Committee) FOMC meetings and find that stock prices increase when the tone is positive. 
\cite{face_2023} analyze videos of FOMC meetings to assess facial expressions and find that negative expressions are priced into financial markets.}

This work is an essential step in a growing literature that introduces novel types of data to econometric analyses \citep[e.g.][]{Lehrer_Tian_2017, wu2015future, compiani2023demand, donaldson2016view}. We show that information contained in photos of home exteriors can capture previously unobservable information about what affects housing prices. The biggest improvements in prediction accuracy come from using these features from multiple encoders. This ultimately improves predictions about house prices by 3\%. In the following section, we describe the data. Section \ref{sec:methods} describes the three distinct methods we use to incorporate information from these images, and then use this to estimate how information about a home's features can predict housing prices. Section \ref{sec:results} summarizes the findings from these three models, and Section \ref{sec:conclusion} concludes with a discussion of the implications for future research.  

\section{Data}
The property dataset used for this research consists of a comprehensive sample of detached homes sold in the City of Toronto between December 2018 and February 2020. Each property in the dataset includes a comprehensive overview of the property's characteristics, including the housing type, number of bedrooms, number of bathrooms, HVAC system, and the quality of the basement and garage finishings, as well as the list and sale price and dates. Sold prices of the properties were spread along a broad spectrum, with the average selling price of roughly \$1.48M (adjusted for inflation to January 2020; logarithmically equal to 14.07). The wide range in sale values, from a minimum of less than \$400,000 to over \$13M, suggests our dataset includes a diverse range of properties.

The sale price varies by the amenities present on the property. The average home in our dataset has 3.34 bedrooms and 3.06 bathrooms, both of which are positively correlated with the sale price of the home. Two-storey homes are the most prevalent, making up 45\% of the properties, closely followed by bungalows, accounting for 33\%. These figures paint a picture of moderately sized, family-oriented living spaces dominating our sample.

Other features like the presence of a finished or walk-out basement, attached garage, central air conditioning or a basement apartment were all positively correlated with the sale price, which is expected from the hedonic literature. We include these control variables in some of our analyses below. 

Finally, each property was geocoded to a neighborhood and postal code to the forward sortation area. Using the address of each property, we searched for the historic listing for each property on a publicly-accessible sale database. From the listing, we saved the primary image, filtering for images that only included the front of the house (excluding interior or aerial photos, or properties where no photos were present). This left a dataset of 6,887 addresses with both image and property characteristic data. 

\section{Methods}\label{sec:methods}

Our methodology involves first using six unique encoders to identify features within images. These encoded features are then used to predict housing prices using three distinct models: OLS, neural networks, or a hybrid of the two. Each model is estimated using each of the six neural network architectures, as well as pairwise combinations of each architecture (for example, all features from the ResNet50 encoder are combined with all the features from the VGG16 encoder), and a ``tout ensemble'' model that uses the features from all individual and pairwise encoder combinations. This approach allows us to determine whether additional information can be obtained by using more than one encoder to classify an image.

\subsection{Turning Images Into Data}\label{sec:encoders}
An encoder is a neural network that transforms an input image into a set of features (an encoding). This is achieved by passing the input through several layers of the network, where each layer captures and represents different aspects of the data. Different encoder models trained on the same input dataset may yield different encodings for the same input image because of the architectural differences in the network layers. These structural variations lead to different feature representations, making some models more suitable for specific types of tasks or data than others. For our purposes, we combine and compare the encodings from four contemporary models: ResNet50 \citep{he2016deep}, VGG16 \citep{simonyan2014very}, Inception Network \citep{szegedy2015going}, and MobileNet \citep{howard2017mobilenets}. 

\begin{figure}
  \includegraphics[width=\linewidth]{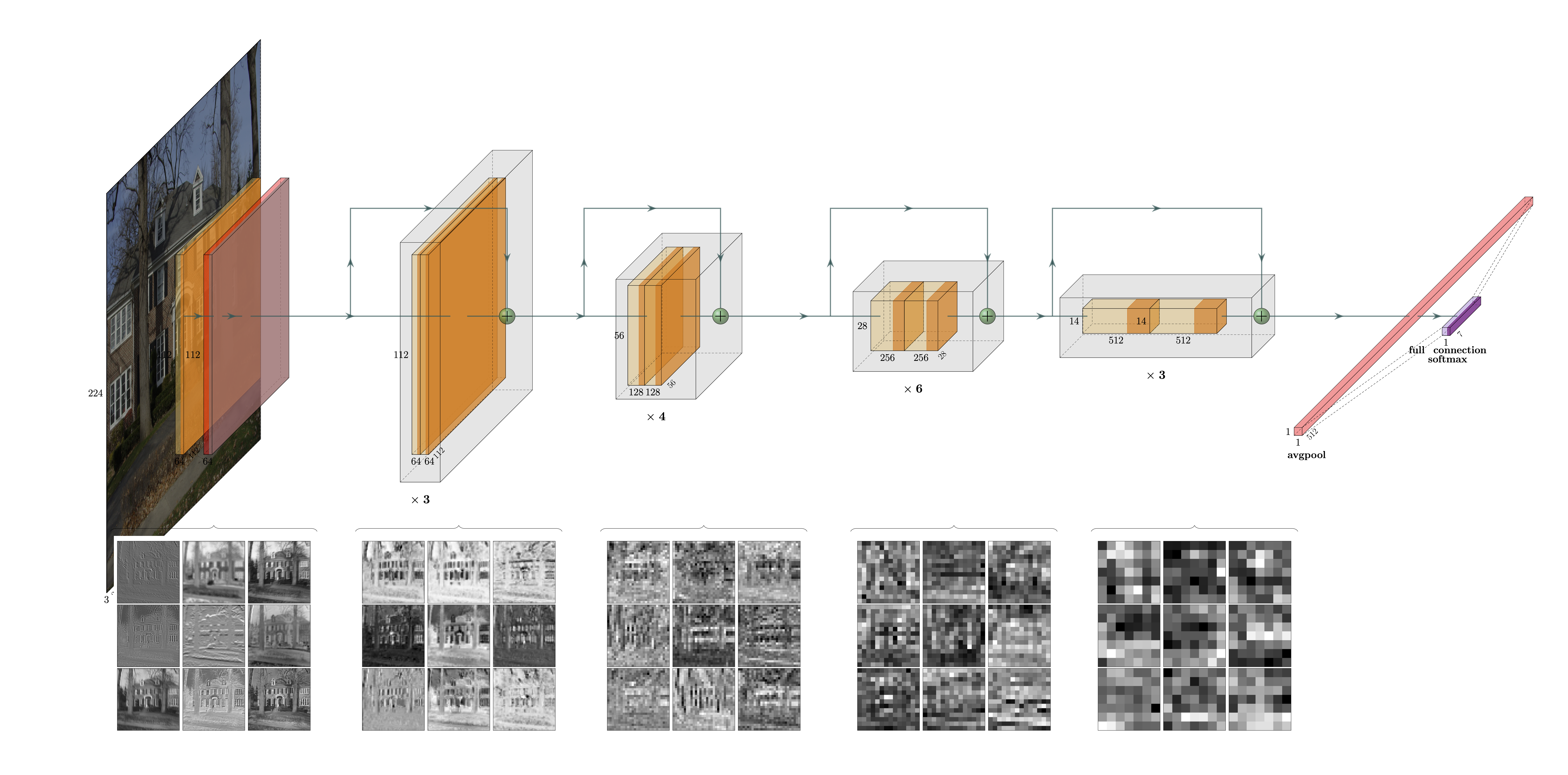}
  \caption{ResNet50 Architecture with Convolution-Layer Predictions }
  \label{fig:resnetArch}
\end{figure}

ResNet50 was a significant innovation in convolutional neural network (CNN) encoder design, introducing a novel solution in the form of residual connections (or ``skip” connections). These connections allow gradients to flow through the network directly by skipping one or more layers. This architectural innovation has enabled the construction of much deeper networks than were previously feasible without a corresponding decrease in performance due to training difficulties. This encoder architecture — and the feature map predicted by its convolutional layers when the “Home Alone” house was provided as an input image — is illustrated in Figure \ref{fig:resnetArch}. 

Like ResNet50, VGG16, Inception Network (also known as GoogleNet), and MobileNet, each uses its own novel architecture to encode an input image and each architecture’s unique approach to processing and extracting features results in different encodings. We used versions of these encoders which had been pre-trained on the ImageNet dataset. ImageNet is a vast collection of over 14 million images from 1,000 different classes, and each model is capable of classifying images into each of those classes and giving a confidence score for that classification \citep{he2016deep, simonyan2014very, szegedy2015going, howard2017mobilenets}. 

In addition to encoding the entire photo using pre-trained models, we also attempted to encode the individual \textit{features} in each image through panoptic segmentation. Panoptic segmentation is a progressive methodology in computer vision that combines semantic and instance segmentation to categorize every pixel in an image into a class (e.g., ``tree”, ``car”, ``building”), and distinguish individual instances of objects within the same class. 

We made use of two commonly used datasets in a pre-trained OneFormer panoptic encoding model: COCO (Common Objects in Context) \citep{lin2014microsoft} and Ade20K \citep{zhou2019semantic}. The COCO dataset has an extensive range of images featuring complex everyday scenes and provides a rich set of contexts for encoding the features in our property photos. It encompasses diverse object categories, both ``stuff” (background features like sky or grass) and ``things” (countable objects like signs or vehicles) that provide rich nuance about the features in each property photo. Similarly, the Ade20K also includes a large number of images spanning diverse settings and object classes. It is part of the MIT Scene Parsing Benchmark and is used extensively in computer vision and panoptic encoding tasks. 

\begin{figure}
  \includegraphics[width=\linewidth]{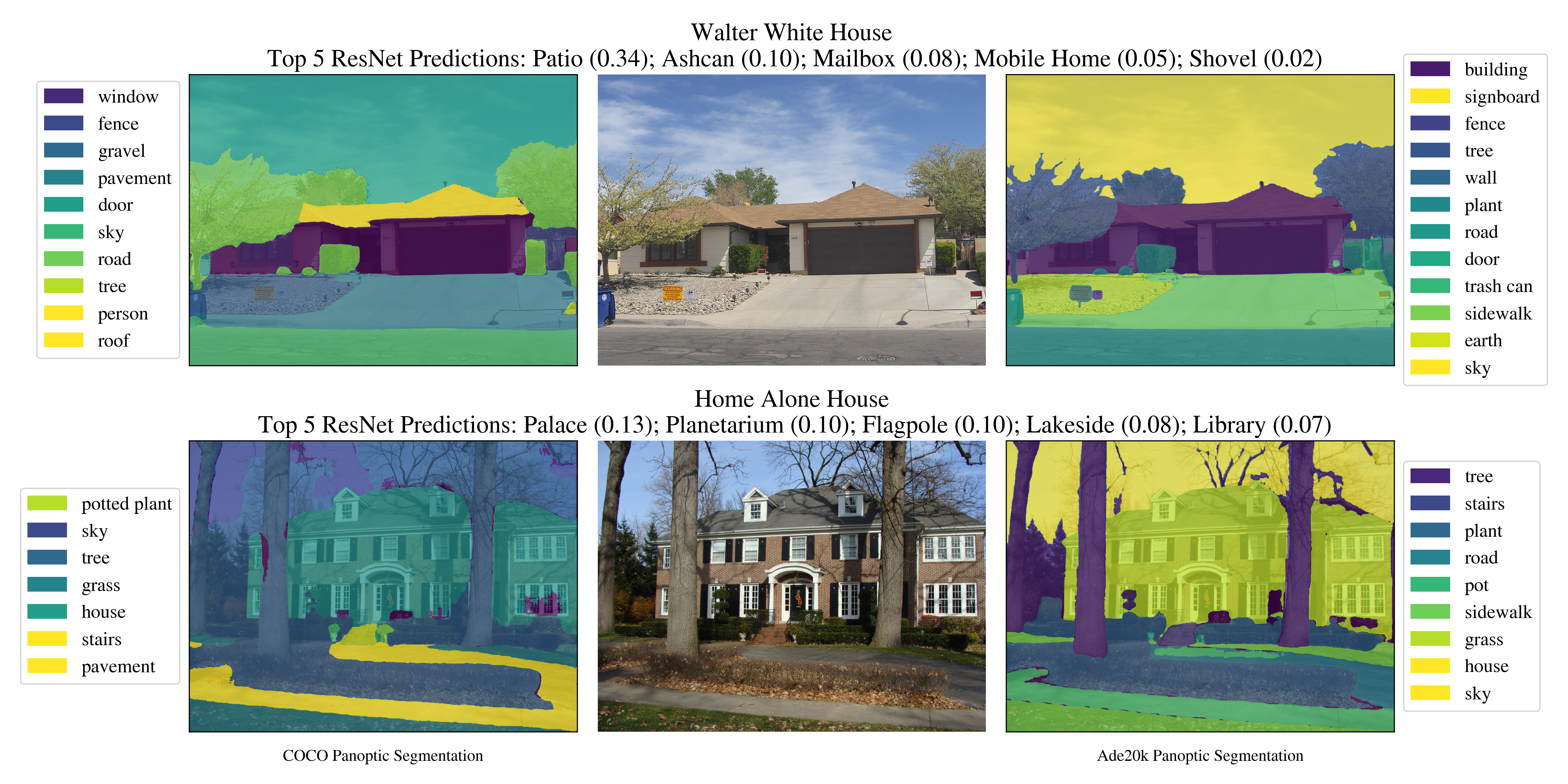}
  \caption{Examples of Encoded Images and Features Within Images}
  \label{fig:EncodingExpample}
\end{figure}

For each property, we encoded how many features from each dataset were identified in the photo, and the proportion of the photo that each feature comprised. Using both datasets allowed us to encode different features for the same property. For example, the COCO dataset has a larger number of classes, but the Ade20k dataset was better able to identify snow. Combining the encodings from both datasets and both the counts and proportion of each feature yields 450 encoded feature variables. Figure \ref{fig:EncodingExpample} illustrates the image-level and panoptic feature encoding of two images of familiar houses from popular culture.  

\subsection{How Image Data are Used to Predict House Prices}

In this paper, we aim to explore several methods for predicting house prices, particularly focusing on whether photos of houses can provide more accurate predictions than merely using observable attributes from real estate listings. Our approach involves comparing different models to determine the most effective strategy.
We specifically investigate the performance of neural networks and OLS, both in isolation and in combination. Our methodology includes using two main types of inputs: the attributes from the Multiple Listing Service (MLS) and the images themselves. For image processing, we employ encoders to categorize the images.

In our first empirical strategy, we estimate OLS regressions to predict house prices. This method involves processing the images through one or more encoders to generate encoded classes. These classes, alongside the MLS attributes, are used as regressors in the OLS model. Given the potential for having more regressors than observations, we use LASSO penalization.  This approach allows us to evaluate the effectiveness of combining traditional listing information with information derived from image data, using a more conventional econometric technique.  A schematic of this strategy is found in \ref{fig:ols}.

\input{modeldiagrams.tex}

Our second strategy is to use neural networks to predict house prices. Each neural network is designed with three intermediate layers. The configuration of these layers is as follows: the first layer contains 128 nodes, the second layer has 64 nodes, and the third layer comprises 32 nodes. The output layer is dedicated to predicting the price, serving as the final goal of the network. The number of input nodes varies significantly, ranging from as few as one or two hundred to over five thousand. This variation depends primarily on which encoder(s) were used and whether the MLS listing attributes are included as inputs.  A schematic of this strategy is found in Figure \ref{fig:neuralnet}.

Our final ``convoluted’’ strategy involves a multi-step process that combines elements of the previous approaches to predict house prices. This process entails several key steps:

\begin{enumerate}
    \item Image Encoding: As above, we run the image through one or more encoders to gather a diverse set of encoded features.
    \item Neural Network Prediction: These encoded features, possibly along with the MLS listing attributes, are then fed into a neural network.  The neural network's task is to predict the house price based on these features.
    \item Combining Predictions with OLS: The predicted price from the neural network, along with the MLS attributes, are then used as regressors in an OLS regression model. This step aims to refine the price prediction by incorporating the neural network's output and the listing attributes into a traditional regression framework.
    \item Final Price Prediction: The OLS model produces the final price prediction. 
\end{enumerate}

This multi-step approach allows us to combine a neural network's ability to handle a great deal of predictors along with the interpretability of OLS. An advantage of this strategy is that it doesn't require LASSO for variable selection. In our sample, the number of regressors (including the predicted price from the neural network and the MLS attributes) is significantly smaller than the number of observations.  A schematic of this strategy is found in Figure \ref{fig:convoluted}.

\subsection{Assessing the Information Content of the Images}

We consider two strategies to assess how much information is contained within the image and how the various models capture that information.  The first approach is to simply regress the actual sales price on the predicted sales price from the model.  To do this analysis we split the sample of houses into both a training set and a testing set.  The neural networks are trained on the training sample and then predictions are made for the testing sample.  Similarly, the OLS models are trained on half of the initial testing dataset.  As a result, we analyze the relationship between predicted and forecasted house prices for a sample of 1,803 homes.  The model used is simply:

\begin{equation}
  \text{price}_h  = \alpha + \beta \widetilde{\text{price}^{att}_h} + \gamma MLS_h.
    \label{eq:price}
\end{equation}

Here $\text{price}_h$ is the actual sales price, in logs, $ \widetilde{\text{price}^{att}_h} $ is the forecasted sales price (also in logs), and  $MLS_h$ is the vector of MLS listing attributes. For this analysis, we restricted attention to the price forecasts from the convoluted neural networks.  We do this mostly as this is the primary methodological innovation in the paper, and there are many ways in which to use both sets of features.

The second strategy is to look at the forecast accuracy of the models and to assess whether adding the image data makes an improvment.  We do this using an 5-fold cross validation approach so that the forecasts are out of sample.  We then compare the Mean Squared Errors (MSEs) for each model type, with each set of inputs.  The MSEs are calculated using the following formula:

\begin{equation}
    MSE_{m,a} = \frac{1}{N}\sum_{h=1}^{N}{ ( \text{price}_h - \widetilde{\text{price}^{m,a}_h}) }.
\end{equation}

Here $m$ indexes a model, and $a$ indexes the set of attributes. We additionally conducted an out-of-sample forecasting exercise in results not shown.

\section{Results with an Application to Housing Price Predictions}\label{sec:results}

In this section we discuss the results of the analysis described above.

\subsection{Do Images Capture Useful Information about Houses?}

Table \ref{tab:ptitle_reg} describes the in-sample results from estimating Equation \eqref{eq:price}. We present the coefficient associated with the predicted price, $\widetilde{\text{price}^{att}_h}$, for each of the encoders, along with the ensemble models that use all encoded features and pairwise feature combinations to predict $\widetilde{\text{price}^{att}_h}$. Panel 1 shows the in-sample relationship that the CNN-predicted prices have on actual sale prices, when the CNN-predicted prices were only based on the features identified by the encoder. We find that for all encoders considered, the predicted prices were significant in explaining the sale price of a home.  Importantly, we show that the predicted prices from the CNNs contain useful information even when the predicted prices were generated using only the features identified by the encoders and not the MLS features. However, the differences between the encoders is considerable.  The $R^2$ can be as low as 0.02 or as high as 0.262.  The highest $R^2$ comes from the ``tout ensemble" model which uses all of the encoders.

\input{p_tilde_regressions}

Panel 2 adds the MLS attributes to the neural networks when predicting $\tilde p$. A similar pattern emerges here, where there is considerable variation in both the $R^2$ and the coefficients, across the encoders.  Again, the model with the highest $R^2$ is the ``tout ensemble'' model.  Interestingly, the coefficient is higher for several of the individual encoders, but their $R^2$ values are considerably lower.  The addition of the attributes to the networks also increases the model fit and the coefficients compared to the estimates based on the image data alone.

The third panel changes how the MLS attributes are used.  Rather than including them as inputs into the neural networks, they are used as regressors in the OLS estimates.  One major change is that the coefficients on predicted prices decrease significantly.  This is to be expected, as there are now many informative covariates in the OLS regressions.  However, the coefficients are all still highly statistically significant even after including the MLS attributes.  Another major change is that the $R^2$ values are much higher, which makes sense given the new covariates.  Again, like in Panel 1 the highest $R^2$ and largest coefficient belong to the ``tout ensemble'' model.

Finally, the fourth panel uses the MLS attributes twice, once as a set of inputs for the neural network, and again as a set of OLS regressors.  The highest overall $R^2$ are obtained from this approach, specifically from the ``tout ensemble'' model. Overall, the change in $R^2$ from Panel 3 is minimal.  However, the variation in coefficients is even more extreme.  No longer are all the coefficients positive and statistically significant.  Now, some of the coefficients are negative and some of those are even statistically significant.  

Taken together these results suggest that the images do indeed contain information that is relevant to the sales price.  Moreover, it appears as though the use of multiple encoders results in superior estimates of the forecasted prices.  Finally, although adding MLS attributes greatly improves the model fit, the images in isolation can have non-trivial explanatory power.

\subsection{Can Data from Images Improve our Housing Price Predictions?}

 To answer this question, we consider whether adding data from the images improves the out-of-sample performance of our three different models. These models incorporate the image data in three distinct ways, leading to three variations of Equation \eqref{eq:price}. 
 We compare the model prediction accuracy to the predictive power of each specification without image data, when only the attributes from the structured data are included in the out-of-sample predictions.

Table \ref{tab:performance_summary_MSE} describes the out-of-sample mean absolute error from each of the different prediction methods. Using just the MLS attributes, we find that the minimum out-of-sample mean squared error from a 5-fold cross-validation is 0.036, 0.126 , 0.036, and 0.044 using penalized OLS, a neural network, and the convoluted model without and with attributes in $\tilde{p}$, respectively. The MSE from penalized OLS alone without any inputs from the images (the ``attribute only'') model is 0.0370. For the two models that could improve upon this, there was a 2.4\%, and a 3.0\% improvement for the penalized OLS specification that made use of the images and features, and the convoluted model with no attributes in $\tilde{p}$, respectively.  The top-performing penalized OLS model was the model using the ``tout ensemble" of encoders.

\input{table_performance_summary_MSE}

The second column of the table considers neural network models, as depicted in Figure \ref{fig:neuralnet}.  On their own, the neural networks have quite high mean squared error rates. The worst convoluted (with or without attributes) model has a better MSE than the best neural network. The neural network with only MLS attributes and no image data has an MSE of 0.1932.  When we consider neural networks based only on the image data the mean of the MSEs is 0.6649, with considerable dispersion between the models.  The best MSE from an image-only neural network
comes from the ``tout ensemble".  Interestingly, when we add MLS attributes to the image data the best MSE improves considerably, and the ``tout ensemble" model is no longer the winner.  Here, the winning model is the pair of MobileNet and Inception encoders.  The ``tout ensemble'' MSE is still respectable at 0.2755, compared to a mean of 0.5857 and a minimum of 0.1259. Conversely, the worst-performing model has a far worse MSE compared to neural networks with no MLS attributes, 5.8715 versus 1.7296.  This corresponds well to the results in Table \ref{tab:ptitle_reg} where some of the worst models have negative coefficients when the MLS attributes are combined with the image data in forming the neural network forecasted price.

The final two columns consider the convoluted models, with and without attributes in the neural network component $\tilde{p}$, as shown in Figure \ref{fig:convoluted}.  Interestingly, the MSEs from the convoluted model without attributes in either the neural network or the OLS stage are better than those from the neural network alone.  This is particularly telling as these models only have an intercept and a slope on $\tilde{p}$ for the forecasting stage.  The mean MSE is 0.2029 compared to the mean neural network MSE of 0.6649. Additionally, if we add MLS attributes to $\tilde{p}$ the MSEs get bigger. This is true both with and without attributes in the regression stage.  These differences are small but noticeable, from means of 0.2029 to 0.2455 for images only, and from means of 0.0367 to 0.0454 for images and attributes.  In all cases, we see that the best-performing MSEs are associated with the ``tout ensemble'' model.  The best performing MSE for convoluted models comes from the ``tout ensemble" model when the MLS attributes are included in the regression but omitted from the neural network. The smallest MSE is 0.0359 when attributes are used only once, versus 0.0443 when used twice.

There are two main takeaways from this analysis.  The first is that using many encoders results in the best MSEs most of the time.  In this setting, this occurred in seven of the eight estimation strategies that we considered.  The second is that using the image and non-image data together seems to minimize the MSE.  For the penalized OLS strategy, using both resulted in the lowest MSE.  The MSE was also lowest for neural networks when using both sets of data as well.  A similar result holds for the convoluted models, but as mentioned, the best result comes from using the MLS data only in the regression stage. 

Our intuition on why multiplier encoders work better than a single encoder is as follows.  Normally when we think about categorizing data we expect that alternative 
classifications may be highly correlated with one another.  For example, recording educational attainment by either years of education, or as the highest level of 
degree or diploma.  Similarly, something like sentiment analysis might result in one person regarding a statement as ``good'' while another person might regard it as ``positive''. 
In these cases, the additional categorizations will not add much value as the classifications are likely to be strongly correlated with each other.

However, in the case of image encoders the classification categories are largely non-overlapping.  This is perhaps surprising, but the number of `things' in the world is obviously unfathomably large so it wouldn't be feasible to assign a probability to each of them.  Thus, one encoder might have `palace' as a category and another has `church'.  Similarly, one encoder might have `aviary' while another has `orchard'.  Thus if we combine multiple encoders together we get different assessments about the likelihood that the images belong to certain categories. Seemingly important for our dataset is that none of the encoders that we used contain a category for `house'.  If they all did, we might be back in an environment where each encoder returned essentially the same information -- that all images were houses. 

\section{Conclusions}\label{sec:conclusion}

These findings demonstrate how deep learning can be used to information from images that is unobservable in traditional data. We apply this to housing, where visual details are particularly relevant since these details influence buyers' perception of a property. However, these subtle visual details are not captured by the characteristics included in traditional housing data. 

Using deep learning we can improve out-of-sample house price predictions using one image of the front of a house. These results are based on an analysis of single-family homes, so the findings may not be directly applicable to other kinds of real estate. However, we believe this paper's main contribution is related to the finding that deep learning and, specifically, the use of multiple encoders can make image data useful in standard econometric analyses. 

An important limitation of the current work is that we are only focused on predicting house prices. We do not attempt to identify any exogenous variations or make any inferences about the causal relationship between the content of these images and housing prices. Other work by \citet{huang2021using} has demonstrated how images from satellites can be used as outcomes in causal analyses, however, we are focused on a different question. Instead, we use data in images as covariates. 

This contributes to a growing area in the economics literature that shows how deep learning can be used to capture information that had previously been unobservable in traditional data sources. Unlike previous work, which has primarily focused on the use of one encoder at a time, we show that by using several deep learning models we can capture more information than would be possible using a single encoder to detect features in image-based data. In our application, using multiple encoders improves the predictive power of all permutations of the models we estimate here. 

Different encoders, such as  Inception Network, ResNet50, MobileNet, and the others used here, are designed with different architectures and are trained on different datasets \citep{szegedy2015going, he2016deep, howard2017mobilenets}. For example, Inception Net's architecture is designed to accommodate visual information that varies in scale and complexity, while VGG architecture is particularly well suited to detect differences in texture and detail \citep{szegedy2015going, kumar2022vi}. This means that by including multiple encoders, we can capture more information from these images. In our setting, this leads to large increases in in-sample model fit and a 3\% improvement in out-of-sample prediction accuracy. Given the growing interest in using images and other unstructured data in economics, we expect these methods will be beneficial to a broad set of applications.

\typeout{} 

\bibliographystyle{cjemod}
\bibliography{bibliography}

\end{document}

%% file: modeldiagrams.tex
\begin{figure}
\centering

\subfloat[Neural Network Model]{
    \input{neuralnet.tex}
    \label{fig:neuralnet}
}

\vspace{1cm} 

\subfloat[OLS Model]{
    \input{ols.tex}
    \label{fig:ols}
}

\vspace{1cm} 

\subfloat[Convoluted Model]{
    \input{convoluted.tex}
    \label{fig:convoluted}
}

\caption{Comparison of Different Models}
\label{fig:models_comparison}
\end{figure}
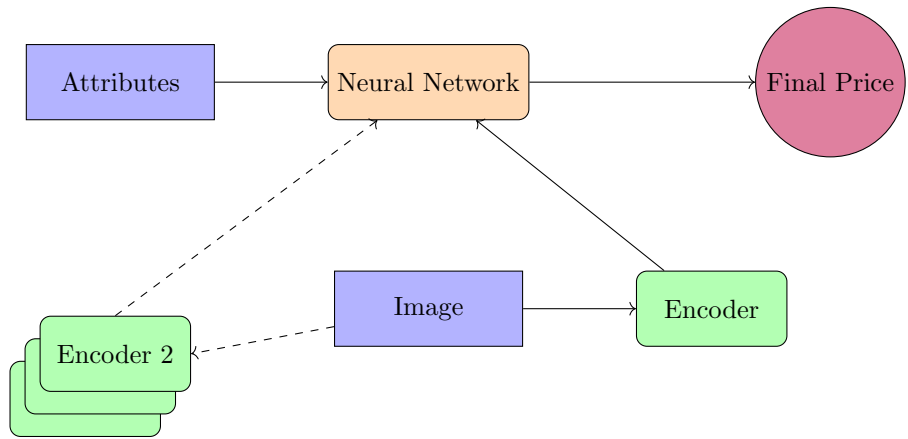
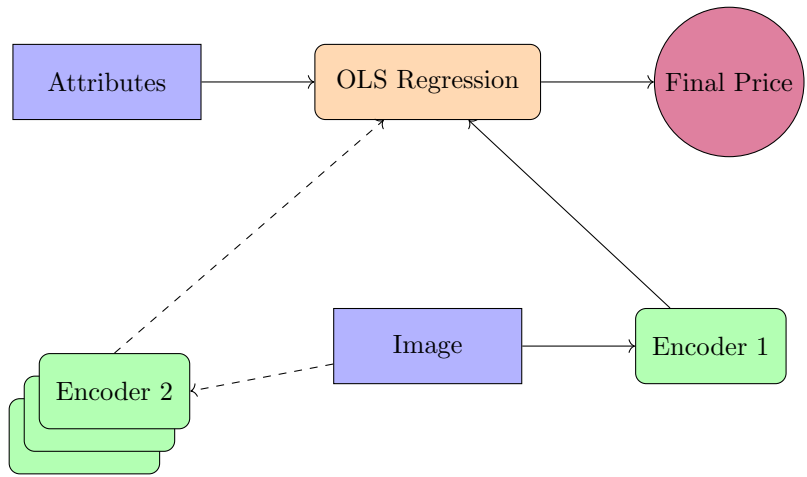
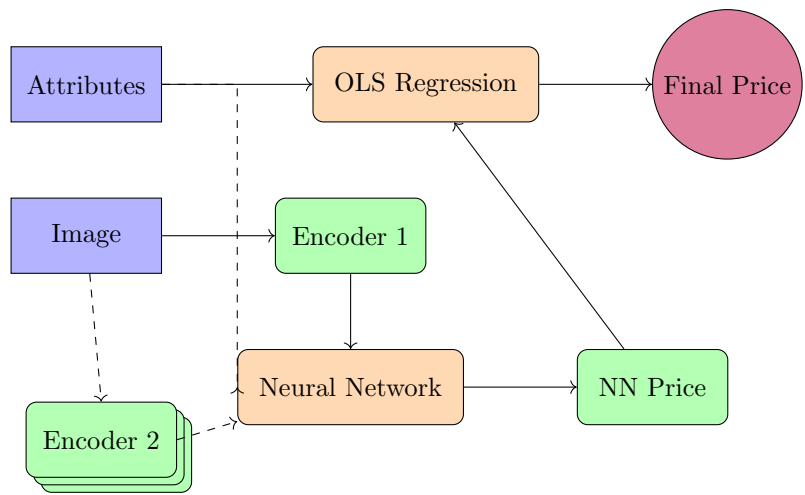

%% file: neuralnet.tex
\centering
\begin{tikzpicture}[node distance=1.5cm,
    input/.style={fill=blue!30, draw, rectangle, minimum width=2.5cm, minimum height=1cm},
    encode/.style={fill=green!30, draw, rectangle, minimum width=2cm, minimum height=1cm, rounded corners},
    dense/.style={fill=orange!30, draw, rectangle, minimum width=2cm, minimum height=1cm, rounded corners},
    final/.style={fill=purple!50, draw, circle, minimum size=1cm},
    cascade/.style={fill=green!30, draw, rectangle, minimum width=2cm, minimum height=1cm, rounded corners, anchor=north west, xshift=-\i*2mm, yshift=\i*2mm}
]

\node[input] (attributes) {Attributes};
\node[dense, right=of attributes] (nn) {Neural Network};
\node[final, right=3cm of nn] (finalprice) {Final Price};

\node[input, below=2cm of nn] (imginput) {Image};
\node[encode, right= of imginput] (encode) {Encoder};

\foreach \i in {4,3,2} {
    \node[cascade, left= of imginput, yshift=-\i*0.5cm] (cascade\i) {Encoder \i};
}

\draw[->] (imginput) -- (encode);
\draw[->] (encode) -- (nn);
\draw[->, dashed] (imginput) -- (cascade2.east);
\draw[->, dashed] (cascade2.north) -- (nn);
\draw[->] (attributes) -- (nn);
\draw[->] (nn) to (finalprice);
\end{tikzpicture}

%% file: ols.tex
\centering
\begin{tikzpicture}[node distance=1.5cm,
    input/.style={fill=blue!30, draw, rectangle, minimum width=2.5cm, minimum height=1cm},
    encode/.style={fill=green!30, draw, rectangle, minimum width=2cm, minimum height=1cm, rounded corners},
    ols/.style={fill=orange!30, draw, rectangle, minimum width=3cm, minimum height=1cm, rounded corners},
    final/.style={fill=purple!50, draw, circle, minimum size=1cm},
    cascade/.style={fill=green!30, draw, rectangle, minimum width=2cm, minimum height=1cm, rounded corners, anchor=north west, xshift=-\i*2mm, yshift=\i*2mm}
]

\node[input] (numinput) {Attributes};
\node[ols, right=of numinput] (ols) {OLS Regression};
\node[final, right=of ols] (finalprice) {Final Price};

\node[input, below=2.5cm of ols] (imginput) {Image};
\node[encode, right=of imginput] (encode) {Encoder 1};

\foreach \i in {4,3,2} {
    \node[cascade, left= of imginput, yshift=-\i*0.5cm] (cascade\i) {Encoder \i};
}

\draw[->] (imginput) -- (encode);
\draw[->, dashed] (imginput) -- (cascade2.east);
\draw[->, dashed] (cascade2.north) -- (ols);
\draw[->] (numinput.east) to (ols.west);
\draw[->] (encode) -- (ols);
\draw[->] (ols) -- (finalprice);

\end{tikzpicture}

%% file: convoluted.tex
    \centering
    \begin{tikzpicture}[node distance=1.5cm,
    input/.style={fill=blue!30, draw, rectangle, minimum width=2cm, minimum height=1cm},
    encode/.style={fill=green!30, draw, rectangle, minimum width=2cm, minimum height=1cm, rounded corners},
    dense/.style={fill=green!30, draw, rectangle, minimum width=2cm, minimum height=1cm, rounded corners},
    ols/.style={fill=orange!30, draw, rectangle, minimum width=3cm, minimum height=1cm, rounded corners},
    final/.style={fill=purple!50, draw, circle, minimum size=1cm},
    cascade/.style={fill=green!30, draw, rectangle, minimum width=2cm, minimum height=1cm, rounded corners, anchor=north west, xshift=\i*1mm, yshift=-\i*1mm}
    ]
    \node[input] (numinput) {Attributes};
    \node[ols, right=2cm of numinput] (ols) {OLS Regression};
    
    \node[input, below=1cm of numinput] (imginput) {Image};
    \node[encode, right=of imginput] (encode) {Encoder 1};
    \node[final, right=of ols] (finalprice) {Final Price};
    
    \foreach \i in {4,3,2} {
        \node[cascade, below= of imginput] (cascade\i) {Encoder \i};
    }

    \node[ols, below=1cm of encode] (nn) {Neural Network};
    \node[encode, right=of nn] (nnprice) {NN Price};
    
    \draw[->] (imginput) -- (encode);
    \draw[->, dashed] (imginput) -- (cascade2.north);
    \draw[->, dashed] (cascade2.east) -- (nn);
    \draw[->] (encode) -- (nn);
    \draw[->] (nn) -- (nnprice);
    \coordinate (midpoint) at ([xshift=1cm]numinput.east);
    \draw[->] (numinput.east) -- (midpoint) |- (ols);
    \draw[->, dashed] (numinput.east) -- (midpoint) |- (nn);
    \draw[->] (nnprice) -- (ols);
    \draw[->] (ols) -- (finalprice);
    \end{tikzpicture}

%% file: p_tilde_regressions.tex
\begin{table}[htbp]
  \centering
  \caption{In-Sample Price Predictions}
    \begin{tabular}{lccccccc}
    \hline
    \hline
         & \multicolumn{7}{c}{Model} \\
          & (1)   & (2)   & (3)   & (4)   & (5)   & (6)   & (7) \\
          & RNF & CC+CT & AC+AT & IC & VGG & MN &  Tout \\
    \hline
    \multicolumn{8}{l}{\textbf{Panel 1: No Attributes in Model, No Attributes in $\tilde{p}$}} \\
     & & & & & & & \\
    $\tilde{p}_{model}$ & 0.223*** & 0.0894*** & 0.129*** & 0.199*** & 0.0526*** & 0.196*** & 0.447*** \\
          & (0.0238) & (0.0206) & (0.0217) & (0.0234) & (0.00917) & (0.0274) & (0.0445) \\
    Attributes &  -     &    -   &    -  &    -   &    -  &   -    & -  \\
    Observations & 1,803 & 1,803 & 1,803 & 1,803 & 1,803 & 1,803 & 1,803 \\
    R-squared & 0.112 & 0.023 & 0.046 & 0.110 & 0.021 & 0.090 & 0.262 \\
    \hline
    \multicolumn{8}{l}{\textbf{Panel 2: No Attributes in Model, Attributes in $\tilde{p}$}} \\
    & & & & & & & \\
    $\tilde{p}_{model}$ & 0.998*** & 0.514*** & 0.966*** & 0.876*** & 0.0901*** & 0.514*** & 0.639*** \\
          & (0.0665) & (0.0817) & (0.0765) & (0.0635) & (0.0177) & (0.0970) & (0.0342) \\
    Attributes &  -     &    -   &    -  &    -   &    -  &   -    & -  \\
    Observations & 1,803 & 1,803 & 1,803 & 1,803 & 1,803 & 1,803 & 1,803 \\
    R-squared & 0.339 & 0.164 & 0.256 & 0.285 & 0.024 & 0.042 & 0.424 \\
    \hline
    \multicolumn{8}{l}{\textbf{Panel 3: Attributes in Model, No Attributes in $\tilde{p}$}} \\
    & & & & & & & \\
    $\tilde{p(no att)}_{model}$ & 0.0309*** & 0.00423 & 0.0133** & 0.0249*** & 0.00925** & 0.0278*** & 0.0684*** \\
          & (0.00735) & (0.00649) & (0.00620) & (0.00729) & (0.00362) & (0.00788) & (0.0144) \\
    Attributes &  \checkmark     &    \checkmark   &    \checkmark   &    \checkmark   &    \checkmark   &   \checkmark    & \checkmark  \\
    Observations & 1,803 & 1,803 & 1,803 & 1,803 & 1,803 & 1,803 & 1,803 \\
    R-squared & 0.872 & 0.871 & 0.871 & 0.872 & 0.871 & 0.872 & 0.875 \\
    \hline
    \multicolumn{8}{l}{\textbf{Panel 4: Attributes in Model, Attributes in $\tilde{p}$}} \\
    & & & & & & & \\
    $\tilde{p(w/att)}_{model}$ & 0.0867*** & 0.0758*** & 0.111*** & 0.0263 & -0.0158** & -0.0405 & 0.118*** \\
          & (0.0282) & (0.0187) & (0.0293) & (0.0331) & (0.00758) & (0.0345) & (0.0165) \\
    Attributes &  \checkmark     &    \checkmark   &    \checkmark   &    \checkmark   &    \checkmark   &   \checkmark    & \checkmark  \\
    Observations & 1,803 & 1,803 & 1,803 & 1,803 & 1,803 & 1,803 & 1,803 \\
    R-squared & 0.872 & 0.873 & 0.873 & 0.871 & 0.871 & 0.871 & 0.879 \\
    \hline
    \hline
    \multicolumn{8}{l}{\footnotesize Robust standard errors in parentheses. Clustered at the 3-digit postal code. *** p$<$0.01, ** p$<$0.05, * p$<$0.1}

    \end{tabular}%
  \label{tab:ptitle_reg}%
\end{table}%

%% file: table_performance_summary_MSE.tex
\begin{table}[]
\centering
\caption{Summary of Model Performance}
\begin{tabular}{lcccc}
                    \hline
                    \hline
                    & \multicolumn{3}{c}{\textbf{Prediction Method}} \\
Inputs              & \multicolumn{1}{c}{Penalized OLS}                                                                 & \multicolumn{1}{c}{Neural Network}                                                   & Convoluted - no & Convoluted -                                                                                               \\
& &  &  attributes in $\tilde{p}$ & attributes in $\tilde{p}$ \\ 
\hline
Attributes          & MSE:  0.0370    &     MSE:     0.1932        &  MSE: -- & MSE: -- \\
\hline
Images              & \begin{tabular}[c]{@{}c@{}}\textbf{Min MSE: 0.1212} \\ Mean MSE: 0.1436 \\ Max MSE: 0.1886 \end{tabular} & \begin{tabular}[c]{@{}c@{}}\textbf{Min MSE: 0.3253}  \\ Mean MSE: 0.6649 \\ Max MSE: 1.7296 \end{tabular} & \multicolumn{1}{c}{\begin{tabular}[c]{@{}c@{}}\textbf{Min MSE: 0.1700}  \\ Mean MSE: 0.2029 \\ Max MSE: 0.2223 \end{tabular}} & \multicolumn{1}{c}{\begin{tabular}[c]{@{}c@{}}\textbf{Min MSE: 0.2046} \\ Mean MSE: 0.2455 \\ Max MSE: 0.2688 \end{tabular}} \\                                                                                                              
\hline
Attributes + Images & \begin{tabular}[c]{@{}c@{}}\textbf{Min MSE: 0.0361} \\ Mean MSE: 0.0374 \\ Max MSE: 0.0393 \end{tabular} & \begin{tabular}[c]{@{}c@{}}Min MSE: 0.1259 \\ Mean MSE: 0.5857  \\ Max MSE: 5.8715 \end{tabular} & \multicolumn{1}{c}{\begin{tabular}[c]{@{}c@{}}\textbf{Min MSE: 0.0359} \\ Mean MSE: 0.0367 \\ Max MSE: 0.0371 \end{tabular}} & \multicolumn{1}{c}{\begin{tabular}[c]{@{}c@{}}\textbf{Min MSE: 0.0443}  \\ Mean MSE: 0.0454 \\ Max MSE: 0.0459 \end{tabular}}\\
\hline
\hline
\end{tabular}
\vskip 6pt
\textbf{Notes:} Bolded numbers represent the MSE associated with the ``tout ensemble" model.  MSE predictions from 1,843 observations using 5-fold cross validation.
\label{tab:performance_summary_MSE}
\end{table}